\pdfoutput=1
\documentclass[10pt, conference, letterpaper]{IEEEtran}
\usepackage{amsmath,amssymb,amsthm}
\usepackage{graphicx}
\usepackage{cite}
\newtheorem{proposition}{Proposition}
\begin{document}
\title{Design and Analysis of Pinching-Antenna Systems With Radially Deployed Waveguides (X-PASS)}
\author{
\IEEEauthorblockN{Wei Jiang}
\IEEEauthorblockA{Intelligent Networking Research Group\\
German Research Center for Artificial Intelligence (DFKI)\\
Kaiserslautern, 67663 Germany}
\and
\IEEEauthorblockN{Hans D. Schotten}
\IEEEauthorblockA{Department of Electrical and Computer Engineering\\
University of Kaiserslautern (RPTU)\\
Kaiserslautern, 67663 Germany}
}
\maketitle
\raggedbottom
\begin{abstract}
Pinching-antenna systems (PASS) reconfigure the wireless channel by positioning dielectric antennas along a waveguide. A single waveguide, however, confines its antennas to one line, so that good coverage of an area has relied on several waveguides or waveguide segments with separate feeds. This paper proposes X-PASS, in which a hub feeds radially deployed waveguide arms, one pinching antenna on each, and drives them by arm selection, aggregation or multiplexing. It models the hub with its insertion losses, derives a unified signal-to-noise ratio expression for these three operating modes, and derives in closed form a radius inside which aggregation outperforms selection. Numerical results validate the analysis and compare the coverage of X-PASS with that of the other layouts at equal transmit power.
\end{abstract}


\section{Introduction}
Pinching-antenna systems (PASS)~\cite{Ref_ding2025pass,Ref_liu2025tutorial} apply a dielectric particle, a ``pinch'', to a waveguide to create a \emph{pinching antenna} (PA) near a user, so that large-scale path loss and line-of-sight blockage become design variables. A single waveguide, however, confines its PAs to one line, and a user beside it keeps its lateral distance wherever the PA moves. Covering an area evenly therefore needs several radiating waveguides or waveguide sections, which the literature obtains by spreading the feed points in space. Multi-waveguide PASS lays parallel waveguides across the area, each with its own feed~\cite{Ref_ding2025pass,Ref_wang2025wdma}, and drives them all at once with one radio-frequency (RF) chain each (waveguide multiplexing) or one at a time through a switch (waveguide switching)~\cite{Ref_zhu2026ws}. Segmented waveguide-enabled PASS (SWAN) cuts a waveguide into short segments, each with its own feed link, and selects, aggregates or multiplexes the segments~\cite{Ref_ouyang2025swan}. Center-fed PASS (C-PASS) feeds a waveguide at its center so that both halves radiate, by power splitting or direction switching~\cite{Ref_gan2026cpasscomm}.

Recent work optimizes the placement of the waveguides, again with a separate feed per waveguide~\cite{Ref_xu2026deployment}, and a recent tutorial lists their number, layout and feeding among the open questions of the field~\cite{Ref_xu2026networkperspective}. These layouts thus either stay on one line, as C-PASS does, or need a feed at a different point of the area for every waveguide or segment. A star of waveguides fed from one hub was used in~\cite{Ref_jiang2026thzpass} to cut the in-guide loss of terahertz links and in~\cite{Ref_jiang2026spass} to schedule antenna movement, but how the hub should drive the waveguides, and what coverage the radial layout gains over the layouts above, have not been analyzed.

To fill this gap, this paper proposes X-PASS, named after its shape: a hub feeds several waveguide \emph{arms} that extend radially to the coverage radius, one PA on each, so that the arms partition the area into sectors. All RF hardware sits at the hub, which drives the arms in one of three operating modes: \emph{arm selection} (AS) switches one RF chain to the arm whose sector holds the user, \emph{arm aggregation} (AA) splits the signal of one chain among all arms with aligned phases, and \emph{arm multiplexing} (AM) gives every arm its own chain. Under AS, X-PASS reduces to these stars, whose terahertz and scheduling aspects are not treated here. The contributions are as follows.
\begin{itemize}
\item \emph{We propose} the X-PASS architecture with its three operating modes, modeled through a hub matrix that carries the switching, splitting and phase settings and the insertion losses of the switch and splitter. One PA per arm also keeps the uplink free of the inter-antenna radiation among several active PAs on one waveguide~\cite{Ref_ouyang2025swan,Ref_wang2025multiport}.
\item \emph{We derive} a unified signal-to-noise ratio (SNR) expression for the three modes and compare them: AM is never worse than the other two, whereas the better single-chain mode depends on the switch and splitter losses and on the user position, with a closed-form radius inside which AA outperforms AS at every azimuth.
\item \emph{We show} that, beyond small areas, X-PASS covers an area from one hub with a higher cell-edge SNR than an end-fed, center-fed or segmented waveguide at equal transmit power, and within a few decibels of parallel waveguides with distributed feeds, whereas under strong attenuation AS falls below C-PASS and SWAN. Numerical results validate the analysis.
\end{itemize}

The rest of this paper is organized as follows. Section~\ref{sec:model} presents the system model, Section~\ref{sec:modes} the operating modes and the unified SNR, Section~\ref{sec:ana} the comparison of the modes, Section~\ref{sec:num} the numerical results, and Section~\ref{sec:concl} concludes.

\section{System Model}\label{sec:model}
X-PASS consists of a hub that feeds $S$ radially deployed waveguide arms, each carrying one PA, and serves $K$ users on the ground. Its architecture, channel, and signal model follow.

\begin{figure}[!t]\centering
\includegraphics[width=0.82\columnwidth]{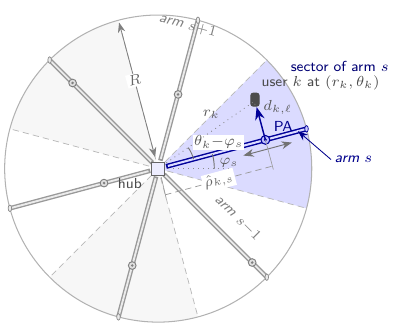}
\caption{Plan view of X-PASS: user $k$ lies in the sector of its assigned arm $s=s_k$, whose PA sits at the projection $\hat\rho_{k,s}$ with lateral offset $d_{k,\ell}$.}\label{fig:geo}\end{figure}
\subsection{X-PASS Architecture}
A hub at $[0,0,h]^{T}$, i.e., at height $h$ above the ground plane, feeds $S\ge2$ separate dielectric waveguides, the \emph{arms}, which extend in the horizontal plane at the azimuths $\varphi_1<\dots<\varphi_S$ to the coverage radius $R$ (Fig.~\ref{fig:geo}).\footnote{More generally, arm $s$ may have its own length $R_s$, with $\rho_s\in[0,R_s]$, so that X-PASS can also cover a square, a rectangle, or an irregular area. This paper treats the circular area, $R_s=R$ for all $s$.} Arm $s\in\{1,\dots,S\}$ carries one movable PA at
\begin{equation}
\boldsymbol{\psi}_s=[\rho_s\cos\varphi_s,\,\rho_s\sin\varphi_s,\,h]^{T},\qquad \rho_s\in[0,R],
\end{equation}
where $\rho_s$ is its distance from the hub along the arm.
The users lie in the circular coverage area of radius $R$. User $k\in\{1,\dots,K\}$ at radius $r_k\le R$ and azimuth $\theta_k$ occupies the ground point $\mathbf{u}_k=[r_k\cos\theta_k,\,r_k\sin\theta_k,\,0]^{T}$ and is assigned to the arm of smallest angular offset,
\begin{equation}\label{eq:assign}
s_k=\arg\max_{s}\,\cos(\theta_k-\varphi_s),
\end{equation}
so that the arms partition the coverage area into $S$ \emph{sectors}.

The distance between user $k$ and the PA at $\boldsymbol{\psi}_s$ is $D_{k,s}\triangleq\|\boldsymbol{\psi}_s-\mathbf{u}_k\|=\sqrt{h^{2}+d_{k,s}^{2}}$, where $d_{k,s}$ is their horizontal distance; both depend on the PA position $\rho_s$, which is not written as an argument. In coordinates, we have
\begin{align}
d_{k,s}^{2}&=(r_k\cos\theta_k-\rho_s\cos\varphi_s)^{2}+(r_k\sin\theta_k-\rho_s\sin\varphi_s)^{2}\nonumber\\
&=r_k^{2}+\rho_s^{2}-2r_k\rho_s\cos(\theta_k-\varphi_s).\label{eq:dist}
\end{align}
Eq.~\eqref{eq:dist} is a convex quadratic in $\rho_s$ with $\partial d_{k,s}^{2}/\partial\rho_s=2[\rho_s-r_k\cos(\theta_k-\varphi_s)]$, so $d_{k,s}$ decreases up to $\rho_s=r_k\cos(\theta_k-\varphi_s)$, the orthogonal projection of user $k$ onto the line of the arm, and increases beyond it. Since the arm exists only for $\rho_s\ge0$, $d_{k,s}$ is minimized over the arm at the \emph{projection}
\begin{equation}\label{eq:proj}
\hat\rho_{k,s}=\max\{0,\,r_k\cos(\theta_k-\varphi_s)\}.
\end{equation}
On an arm with $\cos(\theta_k-\varphi_s)<0$ the projection is the hub and $d_{k,s}=r_k$. The analysis takes equally spaced azimuths, $\varphi_s=\varphi_1+2\pi(s-1)/S$, so that every sector has the half-angle $\pi/S$. On the assigned arm, $s=s_k$, then $\cos(\theta_k-\varphi_s)\ge\cos(\pi/S)\ge0$, so that the projection lies on the arm, and inserting \eqref{eq:proj} into \eqref{eq:dist} gives the minimum of $d_{k,s}$ over the arm, the \emph{lateral offset}
\begin{equation}\label{eq:ell}
\begin{aligned}
d_{k,\ell}&\triangleq r_k\sqrt{1-\cos^{2}(\theta_k-\varphi_s)}\\
&=r_k|\sin(\theta_k-\varphi_s)|\le R\sin\frac{\pi}{S}.
\end{aligned}
\end{equation}

\subsection{Channel Model}
Assuming line-of-sight links, we adopt the free-space model of~\cite{Ref_ding2025pass,Ref_liu2025tutorial}: the channel coefficient between the PA at $\boldsymbol{\psi}_s$ and user $k$ at $\mathbf{u}_k$ is $g_{\mathrm{fs}}(\mathbf{u}_k,\boldsymbol{\psi}_s)=\sqrt{\eta_0}\,e^{-j(2\pi/\lambda)D_{k,s}}/D_{k,s}$, with $\eta_0=(\lambda/(4\pi))^2$ and $\lambda$ the carrier wavelength. Inside arm $s$ the guided mode travels from the feed to the PA over the length $\rho_s$ and accumulates the attenuation and phase~\cite{Ref_ouyang2025swan}
\begin{equation}
g_{\mathrm{wg}}(\rho_s)=10^{-\frac{\alpha_g\rho_s}{20}}\,e^{-j\frac{2\pi\rho_s}{\lambda_g}},\qquad \lambda_g=\lambda/n_{\mathrm{eff}},
\end{equation}
where $n_{\mathrm{eff}}$ is the effective refractive index and $\alpha_g$ the in-waveguide attenuation in dB/m. The normalized amplitude gain from the feed of arm $s$ through its PA to user $k$ is
\begin{equation}\label{eq:amp}
a_{k,s}\triangleq\frac{10^{-\alpha_g\rho_s/20}}{D_{k,s}},
\end{equation}
so that $|g_{\mathrm{wg}}(\rho_s)\,g_{\mathrm{fs}}(\mathbf{u}_k,\boldsymbol{\psi}_s)|=\sqrt{\eta_0}\,a_{k,s}$. The channel vector between the arm feeds and user $k$ is $\mathbf{g}_k=[g_{k,1},\dots,g_{k,S}]^{T}$ with $g_{k,s}=g_{\mathrm{wg}}(\rho_s)\,g_{\mathrm{fs}}(\mathbf{u}_k,\boldsymbol{\psi}_s)$, and $\mathbf{G}=[\mathbf{g}_1,\dots,\mathbf{g}_K]^{T}\in\mathbb{C}^{K\times S}$ stacks the channels of the $K$ users.

\subsection{Signal Model}
\begin{figure}[!t]\centering
\includegraphics[width=\columnwidth]{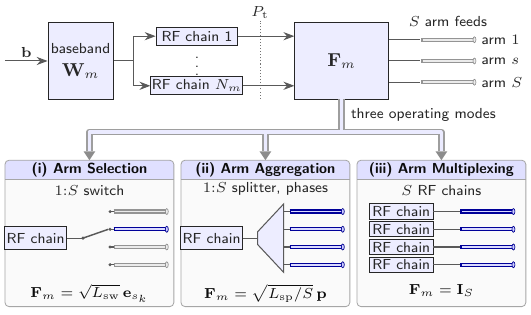}
\caption{Hub of X-PASS: $N_m$ RF chains and the hub matrix $\mathbf{F}_m$ feed the $S$ arms; the three operating modes are realizations of $\mathbf{F}_m$.}\label{fig:modes}\end{figure}

The hub connects $N_m$ RF chains to the $S$ arm feeds through a passive front-end (Fig.~\ref{fig:modes}), described by the \emph{hub matrix} $\mathbf{F}_m\in\mathbb{C}^{S\times N_m}$, with $m$ the operating mode of X-PASS. The matrix maps the chain signals to the feed signals and carries the switching or splitting coefficients, phase settings, and insertion losses; passivity requires $\mathbf{F}_m^{H}\mathbf{F}_m\preceq\mathbf{I}_{N_m}$.
Let $\mathbf{b}\in\mathbb{C}^{B_m}$ denote $B_m$ unit-power data streams, $\mathbb{E}[\mathbf{b}\mathbf{b}^{H}]=\mathbf{I}_{B_m}$, $\mathbf{W}_m\in\mathbb{C}^{N_m\times B_m}$ the precoder with $\|\mathbf{W}_m\|_{\mathrm F}^{2}=P_{\mathrm t}$, where $P_{\mathrm t}$ is the total transmit power of the $N_m$ chains, and $\mathbf{n}\sim\mathcal{CN}(\mathbf{0},\sigma_{\mathrm n}^{2}\mathbf{I}_K)$ the noise at the users. The users receive
\begin{equation}\label{eq:rx}
\mathbf{y}=\mathbf{G}\mathbf{F}_m\mathbf{W}_m\mathbf{b}+\mathbf{n}.
\end{equation}
For ease of analysis, the system operates in time-division duplex, so that the same front-end serves the uplink. With one PA per arm, no captured signal passes another PA, so the uplink is free of inter-antenna radiation~\cite{Ref_ouyang2025swan,Ref_wang2025multiport}. Only the downlink is analyzed, since with the receiver noise at the RF chains the uplink follows by reciprocity.

\section{X-PASS Operating Modes}\label{sec:modes}
This section defines the three operating modes of X-PASS and derives a unified SNR expression for all of them.

\begin{table}[!t]
\centering
\caption{Comparison of the Operating Modes of X-PASS}
\label{tab:prot}
\setlength{\tabcolsep}{3pt}
\footnotesize
\begin{tabular}{lccc}
\hline\hline
 & AS & AA & AM \\
\hline\hline
Hub front-end & $1{:}S$ switch & $1{:}S$ splitter & chain per arm \\
Hub matrix $\mathbf{F}_m$ & $\sqrt{L_{\mathrm{sw}}}\,\mathbf{e}_{s_k}$ & $\sqrt{L_{\mathrm{sp}}/S}\,\mathbf{p}$ & $\mathbf{I}_S$ \\
Hub gain matrix $\boldsymbol{\Gamma}_m$ & $L_{\mathrm{sw}}\mathbf{e}_{s_k}\mathbf{e}_{s_k}^{T}$ & $(L_{\mathrm{sp}}/S)\mathbf{1}\mathbf{1}^{T}$ & $\mathbf{I}_S$ \\[1.5pt]
RF chains $N_m$ & 1 & 1 & $S$ \\
Max.\ data streams $B_m$ & 1 & 1 & $\min\{S,K\}$ \\
PAs positioned per user & 1 & $S$ & $S$ \\
Phase alignment by & -- & phase shifters & precoder \\
\hline\hline
\end{tabular}
\end{table}
\subsection{Arm Selection, Aggregation, and Multiplexing}
Following the three operating protocols of segmented waveguides~\cite{Ref_ouyang2025swan}, we consider three operating modes $m\in\{\mathrm{AS},\mathrm{AA},\mathrm{AM}\}$, shown in Fig.~\ref{fig:modes} and Table~\ref{tab:prot}, each a realization $(N_m,\mathbf{F}_m)$ of the hub, set for the user $k$ being served:
\begin{itemize}
\item \emph{Arm selection:} a $1{:}S$ switch with the insertion loss factor $L_{\mathrm{sw}}(S)$, given in \eqref{eq:loss}, routes the signal of the single chain, $N_m=1$, to the assigned arm $s_k$ of \eqref{eq:assign}. Hence $\mathbf{F}_m=\sqrt{L_{\mathrm{sw}}(S)}\,\mathbf{e}_{s_k}$, where $\mathbf{e}_{s_k}$ is the $s_k^{\mathrm{th}}$ column of $\mathbf{I}_S$, and only arm $s_k$ transmits to user $k$, $|\mathbf{g}_k^{T}\mathbf{F}_m|=\sqrt{\eta_0L_{\mathrm{sw}}(S)}\,a_{k,s_k}$.
\item \emph{Arm aggregation:} a $1{:}S$ splitter with the insertion loss factor $L_{\mathrm{sp}}(S)$, given in \eqref{eq:loss}, divides the signal of the single chain, $N_m=1$, equally among the $S$ arms, and a phase shifter at each splitter output compensates the phase $\angle g_{k,s}=-2\pi(\rho_s/\lambda_g+D_{k,s}/\lambda)$ accumulated in arm $s$ and on its path to user $k$. Hence $\mathbf{F}_m=\sqrt{L_{\mathrm{sp}}(S)/S}\,\mathbf{p}$ with $[\mathbf{p}]_s=e^{-j\angle g_{k,s}}$, and the $S$ arms add coherently at user $k$, $\mathbf{g}_k^{T}\mathbf{F}_m=\sqrt{\eta_0L_{\mathrm{sp}}(S)/S}\,\sum_{s=1}^{S}a_{k,s}$.
\item \emph{Arm multiplexing:} each arm has its own chain, $N_m=S$, and there is neither switch nor splitter. Hence $\mathbf{F}_m=\mathbf{I}_S$, and all $S$ arms transmit to user $k$ through the full channel, $\mathbf{g}_k^{T}\mathbf{F}_m=\mathbf{g}_k^{T}$, whose phases the precoder aligns. With its $S$ chains, AM could also serve up to $\min\{S,K\}$ users in the same resource.
\end{itemize}

The insertion loss factors of the switch and the splitter are
\begin{equation}\label{eq:loss}
L_{\mathrm x}(S)=10^{-\mu_{\mathrm x}\lceil\log_2 S\rceil/10}\le1,\qquad \mathrm x\in\{\mathrm{sw},\mathrm{sp}\},
\end{equation}
where, as for the power dividers of~\cite{Ref_garciarodriguez2016hybrid}, the $1{:}S$ switch and splitter are trees of $1{:}2$ switches and $1{:}2$ splitters with the insertion losses $\mu_{\mathrm{sw}}$ and $\mu_{\mathrm{sp}}$ in dB per device. Since no path from the chain to an arm passes through more than $\lceil\log_2 S\rceil$ devices, \eqref{eq:loss} is exact for $S$ a power of two and, with unequal splitting ratios, conservative otherwise. The factor $L_{\mathrm{sp}}(S)$ excludes the ideal division into $S$ equal parts, which $\mathbf{F}_m$ carries separately, and the phase-shifter loss is neglected.

\subsection{Unified SNR Expression}
AS and AA serve one user with their single chain, and for a fair comparison AM does the same: we assume orthogonal multiple access, e.g., time-division multiple access, with one user per time-frequency resource. Since each user has a single antenna, user $k$ then receives one stream in every mode, $B_m=1$. Row $k$ of \eqref{eq:rx} thus reads $y_k=\mathbf{g}_k^{T}\mathbf{F}_m\mathbf{w}_m b_k+n_k$, with the unit-power data symbol $b_k$ of user $k$ and the precoder $\mathbf{w}_m\in\mathbb{C}^{N_m}$, $\|\mathbf{w}_m\|^{2}=P_{\mathrm t}$, and the SNR of user $k$ is
\begin{equation}\label{eq:snr-dl}
\gamma_{m,k}=\frac{|\mathbf{g}_k^{T}\mathbf{F}_m\mathbf{w}_m|^{2}}{\sigma_{\mathrm n}^{2}}.
\end{equation}
The precoder is maximum-ratio transmission over the effective channel $\mathbf{g}_k^{T}\mathbf{F}_m$, $\mathbf{w}_m=\sqrt{P_{\mathrm t}}\,(\mathbf{F}_m^{T}\mathbf{g}_k)^{*}/\|\mathbf{F}_m^{T}\mathbf{g}_k\|$, which is a scalar of modulus $\sqrt{P_{\mathrm t}}$ in AS and AA and $\sqrt{P_{\mathrm t}}\,\mathbf{g}_k^{*}/\|\mathbf{g}_k\|$ in AM. The hub knows the position of user $k$ in every mode and, in AA and AM, the phases of $\mathbf{g}_k$. Then \eqref{eq:snr-dl} becomes $\gamma_{m,k}=(P_{\mathrm t}/\sigma_{\mathrm n}^{2})\|\mathbf{F}_m^{T}\mathbf{g}_k\|^{2}$, in which the phases of $\mathbf{g}_k$ cancel, so that with the amplitude vector $\mathbf{a}_k\triangleq[a_{k,1},\dots,a_{k,S}]^{T}$
\begin{equation}\label{eq:snr-uni}
\gamma_{m,k}=\gamma_0h^{2}\,\mathbf{a}_k^{T}\boldsymbol{\Gamma}_m\mathbf{a}_k,
\end{equation}
where $\boldsymbol{\Gamma}_m$ is the real \emph{hub gain matrix}
\begin{equation}\label{eq:gamma-m}
\boldsymbol{\Gamma}_m=\begin{cases}
L_{\mathrm{sw}}(S)\,\mathbf{e}_{s_k}\mathbf{e}_{s_k}^{T}, & m=\mathrm{AS},\\
\dfrac{L_{\mathrm{sp}}(S)}{S}\,\mathbf{1}\mathbf{1}^{T}, & m=\mathrm{AA},\\
\mathbf{I}_S, & m=\mathrm{AM},
\end{cases}
\end{equation}
and $\gamma_0\triangleq P_{\mathrm t}\eta_0/(\sigma_{\mathrm n}^{2}h^{2})$ is the SNR of a user on the ground below the hub, served by a PA at the feed through a lossless hub, and the largest SNR that a single arm can deliver, since $a_{k,s}\le1/h$.

\section{Performance Comparison}\label{sec:ana}
This section compares the SNRs of the three modes through \eqref{eq:snr-uni}, first arm multiplexing against the single-chain modes and then arm aggregation against arm selection.

\subsection{Dominance of Arm Multiplexing}
Substituting \eqref{eq:gamma-m} into \eqref{eq:snr-uni} yields the SNR of each mode,
\begin{equation}\label{eq:snr-modes}
\gamma_{m,k}=\gamma_0h^{2}\begin{cases}
L_{\mathrm{sw}}(S)\,a_{k,s_k}^{2}, & m=\mathrm{AS},\\
\dfrac{L_{\mathrm{sp}}(S)}{S}\,\Big(\displaystyle\sum_{s=1}^{S}a_{k,s}\Big)^{2}, & m=\mathrm{AA},\\
\displaystyle\sum_{s=1}^{S}a_{k,s}^{2}, & m=\mathrm{AM}.
\end{cases}
\end{equation}
Hereafter, unless stated otherwise, $\alpha_g=0$ and each PA that carries the signal of user $k$ is at its projection \eqref{eq:proj}, where $a_{k,s}$ is maximal; attenuation moves the optimal position toward the hub~\cite{Ref_xu2026attenuation,Ref_tyrovolas2026performance} and is treated numerically.
\begin{proposition}[SNR advantage of arm multiplexing]\label{prop:bound}
For every user position, we have
\begin{equation}\label{eq:order1}
\begin{aligned}
\gamma_{\mathrm{AM},k}&\ge\frac{\gamma_{\mathrm{AA},k}}{L_{\mathrm{sp}}(S)}\ge\gamma_{\mathrm{AA},k},\\
\gamma_{\mathrm{AM},k}&\ge\frac{\gamma_{\mathrm{AS},k}}{L_{\mathrm{sw}}(S)}\ge\gamma_{\mathrm{AS},k},
\end{aligned}
\end{equation}
that is, AM is never worse than AA and AS, even if their hubs were lossless, $L_{\mathrm{sp}}(S)=L_{\mathrm{sw}}(S)=1$.
\end{proposition}
\begin{IEEEproof}
Dividing \eqref{eq:snr-modes} by $\gamma_0h^{2}a_{k,s_k}^{2}$, the SNR of the assigned arm alone without insertion loss, puts the three modes on one scale,
\begin{equation}\label{eq:norm}
\frac{\gamma_{m,k}}{\gamma_0h^{2}a_{k,s_k}^{2}}=\begin{cases}
L_{\mathrm{sw}}(S), & m=\mathrm{AS},\\
\dfrac{L_{\mathrm{sp}}(S)}{S}\,\Big(\displaystyle\sum_{s=1}^{S}\frac{a_{k,s}}{a_{k,s_k}}\Big)^{2}, & m=\mathrm{AA},\\
\displaystyle\sum_{s=1}^{S}\frac{a_{k,s}^{2}}{a_{k,s_k}^{2}}, & m=\mathrm{AM},
\end{cases}
\end{equation}
in which every normalized gain satisfies $0<a_{k,s}/a_{k,s_k}\le1$, with equality for $s=s_k$, since the assigned arm has the smallest horizontal distance by \eqref{eq:assign} and \eqref{eq:proj} and thus, for $\alpha_g=0$, the largest gain. Hence, expanding the square in the second step and applying the Cauchy--Schwarz inequality in the third,
\begin{equation}\label{eq:nbounds}
1\le\sum_{s=1}^{S}\frac{a_{k,s}^{2}}{a_{k,s_k}^{2}}\le\Big(\sum_{s=1}^{S}\frac{a_{k,s}}{a_{k,s_k}}\Big)^{2}\le S\sum_{s=1}^{S}\frac{a_{k,s}^{2}}{a_{k,s_k}^{2}}\le S^{2},
\end{equation}
where the last two steps hold with equality at the hub, $r_k=0$, since by \eqref{eq:proj} all PAs then sit at the hub, their coupling neglected, and all $a_{k,s}=1/h$. The first two steps hold with equality in the limit far from the hub on the arm axis, where $a_{k,s}\to0$ for $s\ne s_k$. In \eqref{eq:norm}, AS is fixed, $\gamma_{\mathrm{AS},k}/(\gamma_0h^{2}a_{k,s_k}^{2})=L_{\mathrm{sw}}(S)$, whereas AM and AA range as
\begin{equation}\label{eq:ranges}
\begin{aligned}
1&\le\frac{\gamma_{\mathrm{AM},k}}{\gamma_0h^{2}a_{k,s_k}^{2}}\le S,\\
\frac{L_{\mathrm{sp}}(S)}{S}&\le\frac{\gamma_{\mathrm{AA},k}}{\gamma_0h^{2}a_{k,s_k}^{2}}\le SL_{\mathrm{sp}}(S),
\end{aligned}
\end{equation}
with the upper bounds attained at the hub. The AM row of \eqref{eq:norm} and the first step of \eqref{eq:nbounds} give
\begin{equation}\label{eq:am-as}
\gamma_{\mathrm{AM},k}=\gamma_0h^{2}a_{k,s_k}^{2}\sum_{s=1}^{S}\frac{a_{k,s}^{2}}{a_{k,s_k}^{2}}\ge\gamma_0h^{2}a_{k,s_k}^{2}=\frac{\gamma_{\mathrm{AS},k}}{L_{\mathrm{sw}}(S)},
\end{equation}
the last equality being the AS row of \eqref{eq:norm}. The AA row of \eqref{eq:norm} and the third step of \eqref{eq:nbounds} give
\begin{equation}\label{eq:aa-am}
\begin{aligned}
\gamma_{\mathrm{AA},k}&=\frac{L_{\mathrm{sp}}(S)}{S}\,\gamma_0h^{2}a_{k,s_k}^{2}\Big(\sum_{s=1}^{S}\frac{a_{k,s}}{a_{k,s_k}}\Big)^{2}\\
&\le L_{\mathrm{sp}}(S)\,\gamma_0h^{2}a_{k,s_k}^{2}\sum_{s=1}^{S}\frac{a_{k,s}^{2}}{a_{k,s_k}^{2}}=L_{\mathrm{sp}}(S)\,\gamma_{\mathrm{AM},k},
\end{aligned}
\end{equation}
with equality at the hub. Since $L_{\mathrm{sp}}(S),L_{\mathrm{sw}}(S)\le1$ by \eqref{eq:loss}, \eqref{eq:am-as} and \eqref{eq:aa-am} give \eqref{eq:order1}.
\end{IEEEproof}
AM thus outperforms both single-chain modes at every user position, at the price of $S$ chains, and its model has no switch or splitter loss. By \eqref{eq:nbounds}, its margin never exceeds $S/L_{\mathrm{sp}}(S)$ over AA and $S/L_{\mathrm{sw}}(S)$ over AS. Only the upper bounds in \eqref{eq:nbounds} and \eqref{eq:ranges} use the placement at the projections, so that \eqref{eq:order1} extends to any common PA positions and any $\alpha_g$.

\subsection{Comparison of the Single-Chain Modes}
The better single-chain mode depends on the switch and splitter losses and on the user position: AA splits the power among all arms and is better when all PAs are at similar distances from the user, whereas AS feeds only the assigned arm and is better when its PA is much closer than the others.
\begin{proposition}[Arm aggregation versus arm selection]
For every user position, we have
\begin{equation}\label{eq:regimes}
\begin{cases}
\gamma_{\mathrm{AA},k}\ge\gamma_{\mathrm{AS},k}, & L_{\mathrm{sw}}(S)\le L_{\mathrm{sp}}(S)/S,\\
\gamma_{\mathrm{AS},k}\ge\gamma_{\mathrm{AA},k}, & L_{\mathrm{sw}}(S)\ge SL_{\mathrm{sp}}(S),
\end{cases}
\end{equation}
whereas for $L_{\mathrm{sp}}(S)/S<L_{\mathrm{sw}}(S)<SL_{\mathrm{sp}}(S)$ the two modes change order in a sufficiently large coverage area, with AA outperforming AS near the hub at every azimuth,
\begin{equation}\label{eq:aavsas}
\gamma_{\mathrm{AA},k}>\gamma_{\mathrm{AS},k}\quad\text{for }r_k<r^{\star},
\end{equation}
where the guaranteed radius is given by
\begin{equation}\label{eq:rstar}
r^{\star}=h\sqrt{\bigg(\frac{S-1}{\sqrt{SL_{\mathrm{sw}}(S)/L_{\mathrm{sp}}(S)}-1}\bigg)^{2}-1},
\end{equation}
which scales with $h$, grows with $\mu_{\mathrm{sw}}$, and shrinks with $\mu_{\mathrm{sp}}$.
\end{proposition}
\begin{IEEEproof}
The AA and AS rows of \eqref{eq:norm} give
\begin{equation}\label{eq:crit}
\begin{aligned}
\gamma_{\mathrm{AA},k}>\gamma_{\mathrm{AS},k}&\iff\frac{L_{\mathrm{sp}}(S)}{S}\Big(\sum_{s=1}^{S}\frac{a_{k,s}}{a_{k,s_k}}\Big)^{2}>L_{\mathrm{sw}}(S)\\
&\iff\sum_{s=1}^{S}\frac{a_{k,s}}{a_{k,s_k}}>\sqrt{\tfrac{SL_{\mathrm{sw}}(S)}{L_{\mathrm{sp}}(S)}}\triangleq\tau(S),
\end{aligned}
\end{equation}
where both steps, a division by $\gamma_0h^{2}a_{k,s_k}^{2}>0$ and a square root, preserve any order relation, so that \eqref{eq:crit} also holds with $\ge$, $\le$, or $<$ in place of $>$. The threshold $\tau(S)$ is a constant, whereas the sum on the left of \eqref{eq:crit} lies in $[1,S]$ by \eqref{eq:nbounds}.

Hence, for $L_{\mathrm{sw}}(S)\le L_{\mathrm{sp}}(S)/S$, i.e., $\tau(S)\le1$, the threshold and the sum satisfy
\begin{equation}
\tau(S)\le1\le\sum_{s=1}^{S}\frac{a_{k,s}}{a_{k,s_k}},
\end{equation}
so that $\gamma_{\mathrm{AA},k}\ge\gamma_{\mathrm{AS},k}$ at every user position by \eqref{eq:crit}, the first case of \eqref{eq:regimes}. For $L_{\mathrm{sw}}(S)\ge SL_{\mathrm{sp}}(S)$, i.e., $\tau(S)\ge S$, the sum and the threshold satisfy
\begin{equation}
\sum_{s=1}^{S}\frac{a_{k,s}}{a_{k,s_k}}\le S\le\tau(S),
\end{equation}
and, with the inequalities reversed, \eqref{eq:crit} reads
\begin{equation}\label{eq:critrev}
\gamma_{\mathrm{AS},k}>\gamma_{\mathrm{AA},k}\iff\sum_{s=1}^{S}\frac{a_{k,s}}{a_{k,s_k}}<\tau(S),
\end{equation}
so that, by the non-strict form of \eqref{eq:critrev}, $\gamma_{\mathrm{AS},k}\ge\gamma_{\mathrm{AA},k}$ everywhere, the second case.

For $L_{\mathrm{sp}}(S)/S<L_{\mathrm{sw}}(S)<SL_{\mathrm{sp}}(S)$, i.e., $1<\tau(S)<S$, the threshold lies inside the range $[1,S]$ of the sum, and the order changes with the user position, as the two ends of the range show. Since $\alpha_g=0$, \eqref{eq:amp} gives $a_{k,s}=1/D_{k,s}$, so that each ratio in the sum equals $D_{k,s_k}/D_{k,s}$. At the hub, $r_k=0$, every PA sits at the hub by \eqref{eq:proj}, all distances equal $h$, and
\begin{equation}
\sum_{s=1}^{S}\frac{a_{k,s}}{a_{k,s_k}}=\sum_{s=1}^{S}\frac{h}{h}=S>\tau(S),
\end{equation}
so that AA wins at the hub by \eqref{eq:crit}. On the axis of the assigned arm, $\theta_k=\varphi_{s_k}$, its PA sits right above the user, $D_{k,s_k}=h$, whereas every other PA, at the hub or at its projection, is at least $r_k\sin(\min\{2\pi/S,\pi/2\})$ away horizontally, since its arm is at least $2\pi/S$ away in azimuth, so that $D_{k,s}\to\infty$ as $r_k\to\infty$ and
\begin{equation}
\sum_{s=1}^{S}\frac{a_{k,s}}{a_{k,s_k}}=1+\sum_{s\ne s_k}\frac{h}{D_{k,s}}\to1<\tau(S),
\end{equation}
so that AS wins far out on the axis by \eqref{eq:critrev}. Between the hub and the far end of the axis, the sum thus crosses $\tau(S)$: a crossover exists in a sufficiently large coverage area.

A radius inside which AA wins at every azimuth follows from a lower bound on the sum that holds at every user position. Setting $\rho_s=0$ in \eqref{eq:dist} gives $d_{k,s}=r_k$ for a PA at the hub, and \eqref{eq:proj} minimizes $d_{k,s}$ over the arm, so that $d_{k,s}\le r_k$ for every PA at its projection and $a_{k,s}=1/\sqrt{h^{2}+d_{k,s}^{2}}\ge1/\sqrt{h^{2}+r_k^{2}}$. For the assigned arm, $d_{k,s_k}\ge0$ gives $a_{k,s_k}=1/\sqrt{h^{2}+d_{k,s_k}^{2}}\le1/h$, with equality when the user lies on the assigned arm. Splitting off the assigned arm, whose term is $1$, we obtain
\begin{equation}\label{eq:gbound}
\begin{aligned}
\sum_{s=1}^{S}\frac{a_{k,s}}{a_{k,s_k}}&=1+\sum_{s\ne s_k}\frac{a_{k,s}}{a_{k,s_k}}\ge1+(S-1)\,\tfrac{1/\sqrt{h^{2}+r_k^{2}}}{1/h}\\
&=1+\frac{(S-1)\,h}{\sqrt{h^{2}+r_k^{2}}},
\end{aligned}
\end{equation}
whose right side depends on $r_k$ only and decreases from $S$ at $r_k=0$ toward $1$ as $r_k\to\infty$. Setting it equal to $\tau(S)$, $1+(S-1)h/\sqrt{h^{2}+r_k^{2}}=\tau(S)$, and solving for $r_k$ yields
\begin{equation}
\begin{aligned}
\frac{h}{\sqrt{h^{2}+r_k^{2}}}&=\frac{\tau(S)-1}{S-1}\\
\iff r_k&=h\sqrt{\bigg(\frac{S-1}{\tau(S)-1}\bigg)^{2}-1}=r^{\star},
\end{aligned}
\end{equation}
the guaranteed radius \eqref{eq:rstar}, which is real since $1<\tau(S)<S$.

Hence, for $r_k<r^{\star}$, the right side of \eqref{eq:gbound} exceeds $\tau(S)$, since it decreases in $r_k$ and equals $\tau(S)$ only at $r^{\star}$, and
\begin{equation}
\sum_{s=1}^{S}\frac{a_{k,s}}{a_{k,s_k}}\ge1+\frac{(S-1)\,h}{\sqrt{h^{2}+r_k^{2}}}>\tau(S),
\end{equation}
so that $\gamma_{\mathrm{AA},k}>\gamma_{\mathrm{AS},k}$ by \eqref{eq:crit}, which is \eqref{eq:aavsas}.
\end{IEEEproof}
Since \eqref{eq:gbound} is a lower bound on the sum, the radius $r^{\star}$ guarantees the advantage of AA at every azimuth, whereas AA may also win farther out. In summary, for $1<\tau(S)<S$ and in a sufficiently large coverage area,
\begin{equation}
\begin{aligned}
&\gamma_{\mathrm{AM},k}\ge\gamma_{\mathrm{AA},k}>\gamma_{\mathrm{AS},k} &&(r_k<r^{\star}),\\
&\gamma_{\mathrm{AM},k}\ge\gamma_{\mathrm{AS},k}>\gamma_{\mathrm{AA},k} &&(\text{far out on the arm axis}).
\end{aligned}
\end{equation}

\section{Numerical Results}\label{sec:num}
This section validates the analysis and compares X-PASS with the other architectures on their common task, the coverage of a circular area of radius $R$, with all waveguides at the height $h=3$~m. Unless stated otherwise, $S=8$, $\mu_{\mathrm{sw}}=1.5$~dB and $\mu_{\mathrm{sp}}=0.4$~dB per device in \eqref{eq:loss}, and $\alpha_g=0.08$~dB/m as in~\cite{Ref_ouyang2025swan} or a pessimistic $1$~dB/m. Every PA sits where its gain \eqref{eq:amp} over its guide is largest, and all SNRs are relative to $\gamma_0$. Coverage is measured by the cumulative distribution function (CDF) of the SNR over $10^{5}$ uniform user positions, the same for every system, and its 5\textsuperscript{th} percentile serves as the coverage metric, indicating the minimal performance at the cell edge.

\begin{figure}[!t]\centering
\includegraphics[width=0.8\columnwidth]{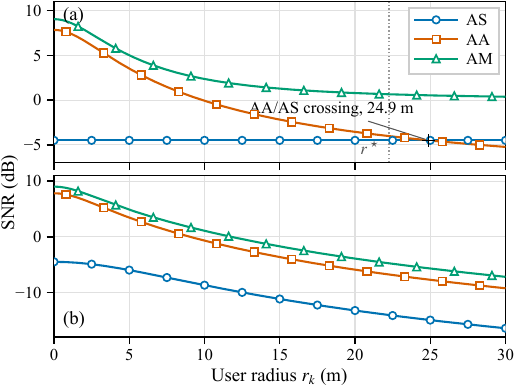}
\caption{SNR of the three modes versus the user radius, $S=8$, $R=30$~m, $\alpha_g=0$: (a) on an arm axis; (b) on a sector boundary. Lines: analysis from \eqref{eq:snr-modes}; markers: simulation from \eqref{eq:snr-dl}; dotted: guaranteed radius $r^{\star}$ from \eqref{eq:rstar}.}\label{fig:modes-r}\end{figure}
\begin{figure}[!t]\centering
\includegraphics[width=0.8\columnwidth]{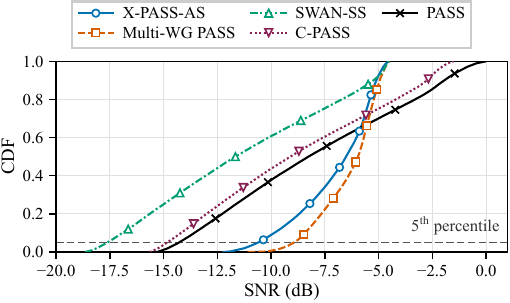}
\caption{CDF of the user SNR, $R=15$~m, $\alpha_g=0.08$~dB/m.}\label{fig:cdf}\end{figure}

Fig.~\ref{fig:modes-r} plots the SNR of the three modes versus the user radius for $\alpha_g=0$ and $R=30$~m, on an arm axis in (a) and on a sector boundary in (b), where the user is nearest to and farthest from its assigned arm. The markers evaluate \eqref{eq:snr-dl} at $f_c=28$~GHz and $n_{\mathrm{eff}}=1.4$ and agree with \eqref{eq:snr-modes} to a relative error of $10^{-15}$. AM is on top everywhere (Proposition~\ref{prop:bound}). At the hub, all PAs are equally near, and AA trails AM only by the splitter loss of $1.2$~dB; farther out, the gap grows, since AA still splits the power equally among PAs at unequal distances. On the axis, AS stays at $10\log_{10}L_{\mathrm{sw}}(8)=-4.5$~dB, since its PA sits right above the user, whereas AA falls from $7.8$~dB at the hub and drops below AS at $24.9$~m, $2.6$~m beyond the guaranteed radius $r^{\star}=22.3$~m. On the boundary, the user lies midway between two arms, so that the sum in \eqref{eq:crit} is at least $2>\tau(8)=1.93$, and AA stays ahead at every radius, by $7.2$~dB at $30$~m. Hence, AS can outperform AA only near the arm axes beyond $r^{\star}$, outside the $15$-m area of Fig.~\ref{fig:cdf}.

\begin{figure}[!t]\centering
\includegraphics[width=0.8\columnwidth]{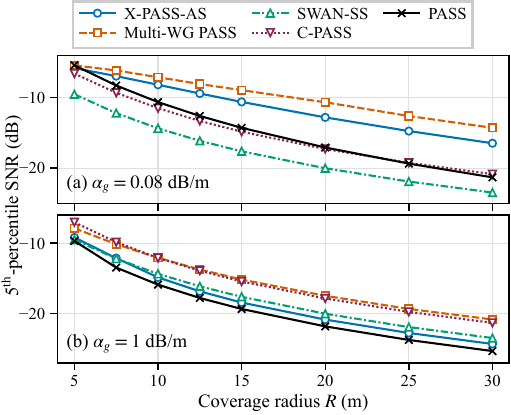}
\caption{5\textsuperscript{th}-percentile SNR versus the coverage radius $R$: (a) $\alpha_g=0.08$~dB/m; (b) $\alpha_g=1$~dB/m.}\label{fig:radius}\end{figure}

Figs.~\ref{fig:cdf} and~\ref{fig:radius} compare X-PASS with four baselines, all at the same power and noise and with lossless feeders to their distributed feed points: \textit{1)~PASS}, the end-fed waveguide along a diameter, length $2R$, with one PA~\cite{Ref_ding2025pass}; \textit{2)~C-PASS} with one input port under direction switching, two half-guides of length $R$ from the center, one PA each, a $1{:}2$ switch~\cite{Ref_gan2026cpasscomm}; \textit{3)~SWAN with segment selection (SWAN-SS)}, the diameter cut into $S$ fed segments, one PA each, a $1{:}S$ switch~\cite{Ref_ouyang2025swan}; \textit{4)~multi-waveguide (multi-WG) PASS}, $S$ parallel waveguides of length $R$, one feed and one PA each~\cite{Ref_ding2025pass}, i.e., the guide length, PAs and chains of X-PASS. Their placement is the best found by a multistart search for the 5\textsuperscript{th}-percentile SNR under waveguide switching~\cite{Ref_zhu2026ws} at $0.08$~dB/m and is kept for the other modes and for $1$~dB/m, with the hub losses of X-PASS.

Fig.~\ref{fig:cdf} plots the CDF of the SNR over the coverage area for $R=15$~m and $0.08$~dB/m. With one chain, X-PASS AS reaches $-10.6$~dB at the 5\textsuperscript{th} percentile, $3.7$, $4.2$ and $7.0$~dB above PASS, C-PASS and SWAN-SS. By \eqref{eq:ell}, its arms keep every user within $R\sin(\pi/S)=5.7$~m of a guide, whereas a diameter leaves the users beside it up to $15$~m away. Multi-WG PASS, with eight feed sites whose straight-line distances to the hub sum to $59$~m, is $1.7$~dB ahead. PASS has a $0.7$~dB higher mean SNR than X-PASS AS, since the users near the diameter have small lateral offsets and no switch loss. Under AA and AM (not shown), X-PASS reaches $-3.6$ and $-1.7$~dB, $0.6$ and $1.1$~dB below multi-WG PASS under aggregation and multiplexing. Hence, at this operating point, X-PASS covers the area from one hub within $2$~dB of the best-found layout with distributed feeds.

Fig.~\ref{fig:radius} plots the 5\textsuperscript{th}-percentile SNR versus the coverage radius for the single-chain schemes. At $0.08$~dB/m, in (a), the lead of X-PASS AS over the single-waveguide architectures is larger in large areas than in small ones. At $R=5$~m, where the lateral offsets are short and the switch loss dominates, AS is still $0.4$~dB below PASS, whereas at $30$~m it leads PASS, C-PASS and SWAN-SS by $4.9$, $4.4$ and $7.0$~dB. Multi-WG PASS stays at most $2.2$~dB ahead. At $1$~dB/m, in (b), the guided run to users beyond about $14$~m costs more than it saves, so their PAs stay at the hub and the link incurs the free-space loss from the hub plus the switch loss. SWAN-SS and multi-WG PASS feed near the rim, and C-PASS has a smaller switch loss. AS then falls up to $3.5$, $3.0$ and $0.8$~dB below multi-WG PASS, C-PASS and SWAN-SS, and from $10$~m on stays ahead of PASS only. AA and AM offset part of this loss, since all PAs of these users sit at the hub and add coherently, coupling neglected. At $R=15$~m, they reach $-6.1$ and $-4.9$~dB, within $0.4$~dB of multi-WG PASS (not shown).

\section{Conclusion}\label{sec:concl}
This paper has proposed X-PASS, in which a hub feeds radially deployed waveguide arms and drives them by arm selection, aggregation or multiplexing, and compared the modes through a unified SNR expression. Arm multiplexing is never worse than the single-chain modes, and aggregation outperforms selection inside a guaranteed radius. Beyond small areas, X-PASS covers an area from one hub with a cell-edge SNR above single-waveguide architectures and close to parallel waveguides with distributed feeds, except for arm selection under strong attenuation. Multi-user operation, several PAs per arm and mixed modes are left for future work.

\bibliography{main}
\end{document}